\documentclass[prd,amsmath,amssymb,showpacs,showkeys,floatfix,
               nofootinbib,eqsecnum,
               preprint,12pt,tightenlines,fleqn
               ]{revtex4-1}
\usepackage{graphicx}

\newcommand{\beq}{\begin{equation}}
\newcommand{\eeq}{\end{equation}}
\newcommand{\bea}{\begin{eqnarray}}
\newcommand{\eea}{\end{eqnarray}}
\newcommand{\bsubeqs}{\begin{subequations}}
\newcommand{\esubeqs}{\end{subequations}}
\newcommand{\nn}{\nonumber}
\newcommand{\q}{\pmb{q}}
\def\k{\kappa}

\begin{document}

\noindent  Phys. Rev. D 92, 025007 (2015) 
\hfill arXiv:1504.01324    
\newline\vspace*{2mm}
%
%

\title{Parton-model calculation of a nonstandard decay process in
isotropic modified Maxwell theory}
\author{J.S. D\'iaz}
\email{jorge.diaz@kit.edu}
\author{F.R. Klinkhamer}
\email{frans.klinkhamer@kit.edu}
\affiliation{Institute for
Theoretical Physics, Karlsruhe Institute of
Technology (KIT), 76128 Karlsruhe, Germany\\}

\begin{abstract}
\noindent
\vspace*{-4mm}\newline
We have performed a calculation in the
parton-model approximation of a nonstandard decay process in
isotropic modified Maxwell theory coupled to standard Dirac theory,
with a single Lorentz-violating parameter $\kappa$ in the photonic sector.
Previous calculations of this process
(vacuum Cherenkov radiation by a proton for the theory with $\kappa>0$)
were performed for pointlike particles,
and an upper bound on $\kappa$ at the $10^{-19}$ level was obtained from experimental data on ultra-high-energy cosmic rays.
The parton-model results change the decay rate by about an order of magnitude but give the same upper bound on $\kappa$
because the threshold energy is unchanged.
The previous point-particle calculation of photon decay into an electron-positron pair for the theory with $\kappa<0$ remains valid,
and experimental results on cosmic gamma rays provide a lower bound on $\kappa$ at the $-10^{-15}$ level.
\end{abstract}


\maketitle

\section{Introduction}

Possible small deviations of Lorentz invariance in the photonic sector
are best bounded by direct experiments in the laboratory (references will be
given later). But it is also possible to obtain tight indirect bounds on
the Lorentz violation by detecting charged or neutral particles with  ultrahigh energies in the Earth's atmosphere, possibly coming from distant astronomical sources.

For these indirect but terrestrial bounds, it is of crucial importance to determine the theoretical decay rates reliably. Up till now, most of these calculations were performed for pointlike particles.
The pointlike calculations are expected to be reliable
for certain decay processes (e.g., photon decay into an electron-positron
pair if kinematically allowed)
but not for other decay processes (e.g., vacuum Cherenkov radiation by
protons or heavy nuclei if kinematically allowed).
These latter processes are better calculated
using the parton-model approximation and the present article
sets out to perform such a calculation for vacuum Cherenkov radiation by a
realistic proton.

\section{Background}

Throughout this article, we use natural units with $\hbar=1=c$ and
the standard Minkowski metric $g_{\mu\nu}(x)=\eta_{\mu\nu}\equiv [\text{diag}(+1,-1,-1,-1)]_{\mu\nu}$.
As the theory considered will be seen to violate Lorentz invariance, we must state clearly what is meant by $c$.  For the particular theory considered,
$c$ is  the maximum attainable velocity of the fermionic particles.

The Lagrange density for modified electrodynamics in the presence of CPT-even Lorentz violation in the photon sector
can be written as follows:
\beq\label{L}
\mathcal{L}_\text{modMaxwell} =
-\frac{1}{4}\,\eta_{\mu\rho}\eta_{\nu\sigma}\, F^{\mu\nu} F^{\rho\sigma}
- \frac{1}{4} (k_F)_{{\mu\nu}\rho\sigma} F^{\mu\nu} F^{\rho\sigma}\, ,
\eeq
in terms of the conventional field strength
tensor $F^{\mu\nu}=\partial^\mu A^\nu-\partial^\nu A^\mu$.
The first term on the right-hand side of \eqref{L}
describes standard Maxwell electrodynamics and the second term corresponds to a dimension-four operator that breaks
Lorentz invariance~\cite{ChadhaNielsen1983,KM2002,BK2004}.
The size of the deviation from exact Lorentz invariance is controlled by the dimensionless coefficient $(k_F)_{{\mu\nu}\rho\sigma}$. The Lagrange density \eqref{L} describes a theory which preserves CPT,
gauge, and coordinate invariance, whereas particle Lorentz invariance is broken.
The coefficient $(k_F)_{{\mu\nu}\rho\sigma}$ has the symmetries of the Riemann tensor and has 20 independent components.
Its double trace can be taken to be zero because it simply modifies the normalization of the Lorentz-invariant part of the Lagrange density \eqref{L}.
In other words, the coefficient $k_F$ is assumed to satisfy
$ (k_F)^{\mu\nu}_{\phantom{{\mu\nu}}{\mu\nu}} = 0, $
which reduces the total number of independent components to 19.

The modified field equations are given by
\beq\label{eom}
\partial^\alpha F_{\alpha\beta} + (k_F)_{{\alpha\beta}{\mu\nu}}\, \partial^\alpha F^{\mu\nu} = 0\, .
\eeq
From \eqref{eom}, it can be seen that
ten of the 19 independent components of the coefficient $(k_F)_{{\alpha\beta}{\mu\nu}}$ produce birefringence in the photon sector,
which can be bounded with remarkable precision using cosmological observations~\cite{CFJ,CarrollField,KM2001}.
For this reason, we focus in the present article on the remaining nine components that produce nonbirefringent effects.
If the ten components generating birefringence are neglected, the coefficient $(k_F)_{{\alpha\beta}{\mu\nu}}$ can be expressed as follows~\cite{BK2004}
\bea\label{kF-nonbirefringent}
(k_F)_{{\alpha\beta}{\mu\nu}} &=& \frac{1}{2} \Big[
 \eta_{\alpha\mu}(k_F)^\lambda_{\phantom{\lambda}\beta\lambda\nu} - \eta_{\alpha\nu}(k_F)^\lambda_{\phantom{\lambda}\beta\lambda\mu}
 \nn\\
 &&\quad
-\eta_{\beta\mu}(k_F)^\lambda_{\phantom{\lambda}\alpha\lambda\nu} + \eta_{\beta\nu}(k_F)^\lambda_{\phantom{\lambda}\alpha\lambda\mu}\Big]\, ,
\eea
where $(k_F)^\lambda_{\phantom{\lambda}\mu\lambda\nu}$
is traceless and symmetric in the indices $\mu$ and $\nu$.
A simple form for this nonbirefringent coefficient can be obtained if we restrict the modified Maxwell theory to describe an
\emph{isotropic} deviation from exact Lorentz invariance.
For this isotropic case, the relevant coefficient in the last expression
of \eqref{L} can be written as \eqref{kF-nonbirefringent} with
\beq\label{kappa}
(k_F)^\lambda_{\phantom{\lambda}\mu\lambda\nu} = \frac{\k}{2}\;\Big[\text{diag}\big(3,1,1,1\big)\Big]_{\mu\nu}\, ,
\eeq
where $\k$ is a short-hand version of $\widetilde{\kappa}_\text{tr}$
used in most of the recent literature.

The isotropic breakdown of Lorentz invariance in the photonic sector
is then controlled by the single coupling constant $\k$
which physically parameterizes a modification of the photon phase velocity.
Modern versions of the Ives--Stilwell experiment~\cite{IvesStilwell} allow for direct laboratory bounds on this parameter down to the
$10^{-10}$ level~\cite{Reinhardt2007,Baynes-etal2012,Michimura-etal2013}.

The parameter $\k$ endows the vacuum with an effective index of refraction
\beq\label{n}
n=\sqrt{(1+\k)/(1-\k)}\, .
\eeq
Writing the photon momentum in the form $q^\mu=(\omega(q),q,0,0)$ for $q\equiv |\q|$, the photon energy and momentum are related by
\beq\label{disp-rel}
\omega(q)  = \frac{1}{n}\,q =  \sqrt{\frac{1-\k}{1+\k}}\;q \, .
\eeq
It will be seen that this modified photon dispersion relation
allows for processes which are kinematically forbidden in the conventional Lorentz-invariant theory.

Lorentz-violating quantum electrodynamics can be constructed by coupling the free photon described by the Lagrange density \eqref{L} to a standard
spin-$\frac12$ Dirac particle $f$ with electric charge $e_{f}$ and inertial mass $M_{f}$,
\bea\label{modQED}
\mathcal{L}_\text{modQED} &=& \mathcal{L}_\text{modMaxwell}+\mathcal{L}_\text{Dirac},
\\[2mm]
\label{L-Dirac}
\mathcal{L}_\text{Dirac} &=& \overline\psi \Big[ \gamma^\mu (i\partial_\mu - e_{f}\,A_\mu) -M_{f}\big]\psi\, .
\eea
The Lorentz-violating operators with the symmetry structure used for the
isotropic modified Maxwell theory can also be moved into the fermion sector
by an appropriate change of spacetime coordinates~\cite{BK2004},
but, for definiteness, we keep the theory as defined
by \eqref{modQED}. The quantum theory corresponding to \eqref{modQED}
has been studied in Ref.~\cite{KS2010}, in particular as regards
microcausality and unitarity.

For negative values of $\k$,
the photon becomes unstable and can produce other particles
by the decay process $\widetilde{\gamma}\to f+\bar f$,
where $f\bar f$ corresponds to a generic pair of electrically charged fermions
and $\widetilde{\gamma}$ denotes the ``photon'' from the isotropic modified Maxwell theory
(the massless spin--1 particle $\widetilde{\gamma}$ has, in particular,
nonstandard polarization vectors, see below).
This process is called photon decay (PhD).
For positive values of $\k$,
photons can be emitted by an electrically charged fermion $f$ and the
process has been called vacuum Cherenkov (VCh) radiation, $f\to f+\widetilde{\gamma}$.

The exact tree-level calculations of the vacuum-Cherenkov and photon
decay processes have been presented in Ref.~\cite{KS2008},
with certain technical details collected in
Ref.~\cite{KaufholdKlinkhamerSchreck2007}.
(The corresponding effects in a CPT-odd photonic theory
coupled to scalar and spinor electrodynamics
have also been studied~\cite{KK2006,KK2007}.)
The absence of experimental evidence for these two nonstandard decay processes has been used in the past to put constraints on $\k$.
Precise measurements of the energy loss of electrons in particle colliders have been considered for determining an indirect laboratory bound on
$|\k|$
of order $10^{-11}$~\cite{Hohensee2009}
and the precise knowledge of the synchrotron loss rate can improve
this bound to the $5 \times 10^{-15}$ level~\cite{Altschul2009}.
But an even tighter two-sided bound follows
from using observations of high-energy gamma rays and cosmic rays in the Earth's atmosphere~\cite{KS2008},
\beq\label{k_bound}
-9\times10^{-16} < \k < 6\times10^{-20}\, ,
\eeq
which holds at the two--$\sigma$ level ($98\,\%$ CL).
Another indirect lower bound~\cite{Altschul2005},
\beq\label{k_qualitative-lower-bound}
\k  \gtrsim -6 \times 10^{-20} \, ,
\eeq
can be obtained by making an appropriate coordinate transformation
to move the isotropic Lorentz violation from the photon sector
into the matter sector (in particular, the electron sector)
and by using the observed synchrotron radiation from high-energy electrons
in the Crab Nebula. Specifically,
the lower bound \eqref{k_qualitative-lower-bound}
results from Eq.~(26) in Ref.~\cite{Altschul2005} and the
following isotropic Lorentz-violating parameters:
$c_{XX}=c_{YY}=c_{ZZ}=-\k$ and $c_{\mu\nu}=0$ otherwise
(note that these $c_{\mu\nu}$  are not the commonly used
traceless-symmetric parameters, cf. Ref.~\cite{Altschul2006}).
The astrophysical bound \eqref{k_qualitative-lower-bound}
is, however, only qualitative,
as Ref.~\cite{Altschul2005} does not give a confidence level.

In most of the literature, the particles involved in these reactions have been considered to be struc\-ture\-less (pointlike).
The description of photon decay into Dirac fermion-antifermion pairs as well as vacuum Cherenkov radiation by a Dirac fermion is appropriate for truly
elementary particles such as electrons.
But, for bounds obtained from particle showers in the Earth's atmosphere,
we need a careful treatment of the internal structure of the primary hadron.

In this work,
we extend the analysis of Ref.~\cite{KS2008} by considering vacuum Cherenkov radiation emitted by a realistic proton at high energies,
relevant for ultra-high-energy cosmic rays.
For completeness, we also discuss
the decay of an energetic astrophysical photon into an electron-positron pair, relevant for observations made with gamma-ray telescopes.
The somewhat more academic case of
photon decay into a proton-antiproton pair is discussed
in App.~\ref{app:Photon-decay-proton-antiproton-pair}.

\section{Vacuum Cherenkov radiation}

For a positive Lorentz-violating parameter $\k$ in the theory \eqref{modQED}, the maximum attainable velocity of a charged particle ($c=1$)
can be larger than the photon phase velocity from \eqref{disp-rel}.
The Cherenkov-like emission of a photon
by a proton of charge $e_{p}\equiv e$
is then already possible in the vacuum.
The modifications introduced by $\k$ to the photon polarization vectors $\tilde\varepsilon^{(\lambda)}_\mu$ are obtained by solving the equation of motion \eqref{eom} for the gauge field $A_\mu$, with
proper normalization factors from the quantum
theory~\cite{KS2010,KaufholdKlinkhamerSchreck2007}.

The averaged squared amplitude for a photon of energy $\omega$
emitted by an electrically charged Dirac fermion with proton mass
$M_{p}\equiv M$ and energy $E$ has the form
\bea\label{<M2>}
\overline{|\mathcal{M}_\text{\,VCh}|^2}  \!\! &=& \!\!
\frac{e^2}{2}\sum_\lambda \sum_{s,s'} \Big|\overline{u}_{s'}(p')\gamma^\mu u_s(p)\,\tilde\varepsilon^{(\lambda)*}_\mu \Big|^2
\nn\\
&=&\!\!
\frac{32\pi\alpha|\k|}{(1+\k)^2}
\bigg[ E(E\!-\omega) \!-\! (1+\k)\, \frac{M^2}{2\k}
\nonumber\\&&
+ \frac{\omega^2}{2(1-\k)}\bigg]\,,\quad
\eea
where $\alpha\equiv e^2/(4\pi)$ is the fine-structure constant.
The total emission rate of Cherenkov photons by a structureless charged fermion is then given by
\beq
\widehat{\Gamma}^\text{\,VCh} =
\frac{1}{4\pi^2}\, \frac{1}{2E}
\int \frac{d^3q}{2\omega} \frac{d^3p'}{2E'} \; \overline{|\mathcal{M}_\text{\,VCh}|^2} \;
\delta^4(p-p'-q)\, ,
\eeq
with  $p'_{0}=E' =\sqrt{|\pmb{p}'|^2+M^2}$ and
$q_{0}=\omega=\omega(|\pmb{q}|)$ as defined by \eqref{disp-rel}.
From now on, the `hat' on $\Gamma$ and similar quantities
signifies that we are considering pointlike particles.
Inserting a factor $\omega$ into the integrand and performing the
necessary phase-space integrals
produces the Cherenkov power radiated by a pointlike ``proton'' of
energy $E$,
\bea\label{P-hat(E)}
\widehat P(E) &=& \frac{\alpha}{12\,\k^3\, E\,\sqrt{E^2-M^2}}
\left(\sqrt{E^2-M^2}- E/n\right)^2
\nn\\
&& \hspace{0.1cm}\times
\Big[2\,E^2\,(2\k^2+4\k+3)
\nn\\
&& \hspace{0.5cm}
-3\,M^2\,(1+\k)(1+2\k)
\nn\\
&& \hspace{0.5cm}
-2\,n\,E\,\sqrt{E^2-M^2}\,(1-\k)\,(4\k+3)
\Big] \,.
\eea
Expression \eqref{P-hat(E)} reduces to Eq. (8) of Ref.~\cite{KS2008}.

If the fermion producing the Cherenkov emission is a realistic proton,
the internal structure becomes relevant at high energies, which is the case for proton primaries in energetic cosmic rays.
Considering the proton as a composite particle,
the Cherenkov photon can be taken as emitted by the charged partons (quarks) in the proton rather than by the proton as a whole (Fig.~\ref{fig:FD-VCh}).  Kinematically, the Cherenkov emission
can occur when a proton of mass $M$ has an energy above the threshold
\beq\label{E_th}
E_\text{th}=M\sqrt{\frac{1+\k}{2\k}}\, .
\eeq
This threshold energy arises from energy-momentum conservation and is independent of the internal structure of the proton.

In this composite-proton description,
Cherenkov radiation corresponds
to the process $p \to X+\widetilde{\gamma}$, where $X$ stands for
any hadronic final state. In fact, the photon emission and the
corresponding parton recoil can lead to a hadronic final state
different from the parent proton. For the description of this
process, we adopt the averaged squared amplitude \eqref{<M2>} for
the emission of a photon by a charged parton.
For a proton with energy $E>E_\text{th}$,
the Cherenkov emission rate produced by a charged parton carrying a fraction $x$ of the proton momentum takes the following form  at tree level:
\bea\label{VCh-emission-rate}
\frac{d^{2}\widehat{\Gamma}_i^\text{\,VCh}}{dx\,d\omega} &=&
\frac{\alpha \, e^2_i}{E\sqrt{E^2-M^2}}
\bigg[\frac{2\k E}{1-\k^2}\, \bigg( E - \frac{\omega}{x} \bigg)
\nn\\
&&
-\frac{M^2}{1-\k}
+ \frac{\k}{(1-\k^2)(1-\k)}\,\frac{\omega^2}{x^2} \bigg]\, ,
\eea
where the index $i$ denotes the parton flavor and
$e_i$ is the parton charge in units of the proton charge $e$.

Notice that expression
\eqref{VCh-emission-rate} allows for the identification of the Cherenkov angle. The obtained Cherenkov angle then includes the classical Huygens term ($\propto\omega^0$)~\cite{Cherenkov1,Vavilov,Cherenkov2,FrankTamm},
the linear quantum correction ($\propto\omega^1$)~\cite{Ginzburg1940,Cox1944,JauchWatson,Ginzburg1996},
and the quadratic quantum correction ($\propto\omega^2$) which arises due to the fermionic nature of the radiating particle~\cite{KK2007}.
Direct calculation shows that, even though the index of refraction is nondispersive,
the energy of the radiated photon has a cutoff given by
\beq\label{w_max}
\omega_\text{max} = x\,\bigg(\frac{1-\k}{\k}\bigg)
 \bigg[ \sqrt\frac{1+\k}{1-\k}\; \sqrt{E^2-M^2} -  E \bigg]\, .
\eeq
Classically, the radiated photon energy is unbounded, leading to a
divergent radiated power. The correct implementation of
energy-momentum conservation relies on the photon's
corpuscular nature, giving rise to a recoil of the fermion due to
the photon emission~\cite{Ginzburg1940,Cox1944,JauchWatson,Ginzburg1996,%
Altschul2008}.
Similarly, the expression for $\omega_\text{max}$ in
\eqref{w_max} is a consequence of the quantum-field-theoretic
treatment of the fermion and photon~\cite{KK2007}.

The expression \eqref{w_max}
of the cutoff energy for the composite proton implies that the Cherenkov emission becomes suppressed for small $x$.
This suggests that the quark sea is irrelevant for Cherenkov radiation and
that  the process is dominated by the valence quarks.
For the same reason, the decay process near threshold
will primarily proceed by
elastic emission of photons, that is, without modifying the structure
of the parent hadron.
Remark that, even though the emitted photon can have ultrahigh energies, the momentum transfer from the primary proton is suppressed by $\k$,
\beq\label{Q2}
Q^2 = -q^2 = \frac{2\,\k\,\omega^2}{1-\k}
\leq \frac{2\,\k}{1-\k}\,\omega_\text{max}^2 =
\text{O}(\k\,x^2\, E^2)\, .
\eeq
The quantity \eqref{Q2} corresponds to the
so-called effective mass square, discussed extensively in Secs.~2 and 4
of Ref.~\cite{KK2006} for various Lorentz-violating decays.
For $E\gg E_\text{th}$, $Q^2$ can be larger than
$M^2$ and the photon emission can be accompanied by the creation of
additional hadrons (baryon number conservation can be assumed to hold true).

The total radiated power by the composite proton
($p \to X+\widetilde{\gamma}$)
is now obtained by folding the parton distribution function $f_i(x,\,Q^2)$
with the power radiated by the corresponding parton,
\beq\label{P(E)}
P(E) = \sum_i \int_0^1dx \int_0^{\omega_\text{max}}d\omega\,
f_i(x,\,Q^2)
\,\omega \; \frac{d^{2}\widehat{\Gamma}_i^\text{\,VCh}}{dx\,d\omega}\, ,
\eeq
where \eqref{VCh-emission-rate} is to be used in the integrand
and $\omega_\text{max}$ and $Q^2$ are given by \eqref{w_max} and
\eqref{Q2}, respectively.
The sum in \eqref{P(E)} contains contributions from all charged
partons in the proton.
Recall that $f_i(x,\,Q^2)$ corresponds to the probability density that a charged parton of flavor $i$ carries a fraction $x$ of the longitudinal momentum, where $Q^2$ parameterizes the resolution scale
at which the parent hadron is probed.
The radiated power \eqref{P(E)} includes all possible hadronic final states.
In fact, expression \eqref{P(E)} is analogous to the parton-model result
for the total cross section of deep-inelastic scattering
$e^{-}+ p \to e^{-}+X$ (cf. Sec.~17.3 of Ref.~\cite{PeskinSchroeder1995}).
From \eqref{P(E)}, the characteristic Cherenkov radiation length is given by
\beq\label{l-VCh}
l_\text{\,VCh}(E)\equiv c\,E/P(E)\, ,
\eeq
with the velocity $c$ temporarily restored.

Libraries with global fits from experimental data as well as the evolution to different energy scales of the parton distribution functions
(PDFs) of the proton are publicly available~\cite{LHAPDF} and
can be used to numerically perform the integrations in \eqref{P(E)}.
With the PDFs from the CTEQ collaboration~\cite{CT10},
the Cherenkov power radiated by a proton is shown in Fig.~\ref{fig:VCh-Power}. In addition, we give in Fig.~\ref{fig:VCh-Power}
the power radiated by a neutron,
which can also be obtained  approximately by the isospin transformation $u\leftrightarrow d$ on the proton PDFs and parton charges.
(Incidentally, the finite lifetime of the neutron
due to $\beta$--decay is irrelevant for the present discussion.)
The ``structureless proton'' in Fig.~\ref{fig:VCh-Power} corresponds to the limit in which the proton is approximated by a pointlike Dirac fermion~\cite{KS2008}. The corresponding Cherenkov radiation lengths are shown in Fig.~\ref{fig:VCh-radiation-length}.
We emphasize that, in our parton-model calculations,
the effective mass square \eqref{Q2}
controls the energy scale for the PDFs rather than the energy of the
incoming proton, as used in the analysis of Ref.~\cite{GagnonMoore}.

Figure~\ref{fig:VCh-Power} shows that the energy loss characterized by the radiated power is lower for the proton than the structureless fermion.
This reduction can be understood as the combination of several effects.
On average, charged partons carry only half of the proton energy,
while the other half is carried by gluons. Furthermore,
the functional form of the power radiated by the proton is similar to that of the structureless fermion folded with the PDFs of each charged parton and with an overall factor $x^2$, where
the momentum fraction $xf_u(x,\,Q^2)$ peaks around $x=0.2$
for relatively low momentum transfer.
These effects give a total suppression factor of approximately
$10$, in agreement with the numerical result shown in Fig.~\ref{fig:VCh-Power}.
This suppression factor slowly increases for higher energies as the $u$-quark sea spreads the momentum fraction to lower values of $x$.
For completeness, the photon emission spectrum
is also given in  Fig.~\ref{fig:VCh-photon-spectrum}.

The detection of a cosmic-ray primary with energy $E_\text{prim}$ implies the condition
\beq\label{E-prim-condition}
E_\text{prim}<E_\text{th}\, ,
\eeq
with $E_\text{th}$ from \eqref{E_th} for $M=M_\text{prim}$. Namely,
if \eqref{E-prim-condition} would not hold, the primary would have lost most of its energy on the journey through space and the Earth's atmosphere.
The validity of the condition $E_\text{prim}<E_\text{th}$
is independent of the astrophysical processes involved in the creation and acceleration of the cosmic-ray primary.
In fact, we can focus on the path through the Earth's atmosphere
(with a height scale of order 10 km), where the particle track can be observed directly (for example, by the fluorescence detectors of
the Pierre Auger Observatory~\cite{Auger2009}
or by gamma-ray telescopes~\cite{Weekes2003}).
The only assumption is that the proton propagation distance $d$
(of the order of 1 km or more) is
significantly larger than the characteristic Cherenkov radiation length,
$d \gg \widehat{l}_\text{\,VCh}$.
The upper bound \eqref{k_bound} on the parameter $\k$ was obtained
from \eqref{E-prim-condition}
by using the detection of a 212-EeV cosmic ray by the Pierre Auger Observatory~\cite{Auger2006}
and by assuming a structureless iron nucleus with
$M_\text{prim}=52\;\text{GeV}$~\cite{KS2008}.
For a primary structureless proton with mass
$M=0.94\;\text{GeV}$, the upper bound in \eqref{k_bound} would be reduced
by a factor $(0.94/52)^2$ to a value of $2 \times 10^{-23}$.

The reduced power radiated by a realistic proton or nucleus with respect to the struc\-ture\-less-fermion result
indicates that this realistic particle will travel over a somewhat larger distance before efficiently losing energy in the form of vacuum Cherenkov radiation.
As discussed above, this extra distance is about one order of magnitude more than the structureless case.
Since the characteristic distance for the photon emission by a structureless fermion is only a fraction of a meter (Fig.~\ref{fig:VCh-radiation-length}),
the assumption of travel distances being larger than the decay length ($d\gg l_\text{\,VCh}$) is completely justified, even for the realistic proton primary described in this work
(for a nucleus $N$, we simply scale the proton results with
$e \to Z_{N}\, e$ and  $M \to M_{N}$).
These remarks imply that the upper bound on the parameter $\k$ as given in Ref.~\cite{KS2008}, based on the threshold condition \eqref{E-prim-condition}, also holds for a realistic proton or nucleus with internal structure taken into account.

\section{Photon Decay}

For a negative Lorentz-violating parameter $\k$ in the theory \eqref{modQED}, the photon becomes unstable and the production of a pair of electrically charged fermions is kinematically allowed at sufficiently high
energies (Fig.~\ref{fig:FD-PhD}).
The averaged squared amplitude for photon decay equals the averaged squared amplitude for vacuum Cherenkov radiation \eqref{<M2>},
\bea
\overline{|\mathcal{M}_\text{\,PhD}|^2} &=&
\frac{e_{f}^2}{2} \sum_\lambda \sum_{s,s'} \Big|\overline{u}_{s}(p)\gamma^\mu v_{s'}(p')\,\tilde\varepsilon^{(\lambda)}_\mu \Big|^2
\nonumber\\
&=& \overline{|\mathcal{M}_\text{\,VCh}|^2}\, .
\eea
The photon decay rate into a fermion-antifermion pair then takes
the form
\beq
\widehat{\Gamma}^\text{\,PhD} =
\frac{1}{4\pi^2}\, \frac{1}{2\omega}
\int \frac{d^3p}{2E} \frac{d^3p'}{2E'} \;\overline{|\mathcal{M}_\text{\,PhD}|^2} \;
\delta^4(q-p-p')\, ,
\eeq
with $p_{0}=E=\sqrt{|\pmb{p}|^2+M_{f}^2}$
and similarly for $p^\prime_{0}=E^\prime$.
Energy-momentum conservation
implies that  pair production
occurs for photon energies above the threshold
\beq\label{w_th}
\omega_\text{th}=2M_{f}\;\sqrt{\frac{1-\k}{-2\k}}\, ,
\eeq
where $M_{f}$ is the equal mass of the fermion and the antifermion
(note that $CPT$ and $C$ are exact symmetries
of the theory considered). Above this threshold,
the decay of a photon of energy $\omega$ produces a fermion with energy $E$ in the range $[E_-,E_+]$,
with
\beq
E_\pm = \frac{\omega}{2}\left[ 1 \pm n\; \sqrt{1-\frac{\omega_\text{th}^2}{\omega^2}} \, \right]\, ,
\eeq
where $n$ is the index of refraction \eqref{n}.
The decay constant is obtained by integrating over the
available fermion energies,
\bea\label{int-dE-dGammahatdE}
\widehat{\Gamma}^\text{\,PhD} &=&
\int_{E_-}^{E_+} \!\!dE\; \frac{d\widehat{\Gamma}^\text{\,PhD}}{dE}\, .
\eea

Specializing to the case of an electron-positron pair
(with charges $e_{f} = \pm e$ and mass $M_{f}=M_{e}\equiv m$) and
performing the energy integral in \eqref{int-dE-dGammahatdE} then gives
\bea
\widehat{\Gamma}^\text{\,PhD}(\omega)
&=&
\frac{\alpha}{3} \,\frac{-\k}{1-\k^2}\;\omega\;
\sqrt{ 1-\omega_\text{th}^2/\omega^2 }\;
\bigg[2+ \omega_\text{th}^2/\omega^2  \bigg]\,,
\qquad
\label{Gamma-hat-PhD}
\eea
where $\alpha\equiv e^2/(4\pi)$ is the fine-structure constant
and $\omega_\text{th}$ is given by \eqref{w_th}
in terms of the electron mass $m$.
Expression \eqref{Gamma-hat-PhD} is equivalent
to Eq. (11) of Ref.~\cite{KS2008}.
The characteristic decay length determined by the photon lifetime
is defined as follows:
\beq\label{l-hat-PhD}
\widehat{l}_\text{\,PhD}(\omega) \equiv c/\widehat{\Gamma}^\text{\,PhD}(\omega)\, ,
\eeq
with the velocity $c$ temporarily restored.

Similar to the case of vacuum Cherenkov radiation,
the photon momentum transfer is suppressed by the parameter $\k$,
which implies that $Q^2=q^2$ is significant only for large values of the photon energy.  In fact, expression \eqref{Q2} gives
\beq\label{Q2-PhD}
Q^2 = \big(2\,m\,\omega/\omega_\text{th}\big)^2\,,
\eeq
with $\omega_\text{th}$ given by \eqref{w_th} in terms of the electron mass
$m$. Again, the quantity \eqref{Q2-PhD} plays the role of
the effective mass square entering
a Lorentz-violating decay~\cite{KK2006}.

The energy scale entering $Q^2$ from \eqref{Q2-PhD} is given by
the mass $2m$ of the electron-positron pair for photon energies close to the threshold.
At these low momentum transfers, the electron is a point particle to very high precision: compositeness bounds
for the electron are at multi-$\text{TeV}$ energy scales
(cf. Ref.~\cite{PDG2014}, pp. 1631--1636).
In short, the point-particle calculation for photon decay
into an electron-positron pair is completely reliable.
The results are shown in Figs.~\ref{fig:PhD-Gamma} and \ref{fig:PhD-decay-length}.
For completeness, the electron emission spectrum is also given in
Fig.~\ref{fig:PhD-electron-spectrum}.

Following the same reasoning as for the case of vacuum Cherenkov radiation,
the observation of a primary photon with energy $\omega_{\gamma,\,\text{prim}}$ implies the condition
\beq\label{omega-prim-condition}
\omega_{\gamma,\,\text{prim}}<\omega_\text{th}\, ,
\eeq
with $\omega_\text{th}$ given by \eqref{w_th} in terms of the electron mass
$m= 511\;\text{keV}$.
Namely, if \eqref{omega-prim-condition} would not hold,
the photon would have decayed along the journey through space and the Earth's atmosphere. Once again,
the validity of this condition is independent of the astrophysical processes involved in the creation of these gamma-ray photons and the only assumption
is that the photons have traveled a distance $d$ larger than the characteristic decay length \eqref{l-hat-PhD}.
The lower bound \eqref{k_bound} on the parameter $\k$ was obtained
from \eqref{omega-prim-condition}
by using the detection of 30-$\text{TeV}$ gamma rays by the High Energy Stereoscopic System (HESS)~\cite{HESS2007}
and by assuming the decay of photons into electron-positron pairs~\cite{KS2008}. In this case,
the decay length of a 30-$\text{TeV}$ photon would be a few millimeters
(Fig.~\ref{fig:PhD-decay-length}),
so that the assumption $d \gg \widehat{l}_\text{\,PhD}$ is
unquestionably valid.

\section{Summary}

In this article, we have studied two nonstandard decay process in CPT-even Lorentz-violating quantum electrodynamics.
In particular, we have considered the effects of an isotropic nonbirefringent modification leading to vacuum Cherenkov radiation by a proton and
photon decay into an electron-positron pair.
Previous studies~\cite{KS2008}
assumed structureless particles for both processes.

For the first process, vacuum Cherenkov radiation by a proton in the
theory \eqref{modQED} with $\k >0$, we have now
also performed a calculation in the framework of the parton model (higher-order QCD corrections have not been considered).
The Cherenkov power is found to be reduced by approximately one order of magnitude
but the threshold energy remains unchanged compared to the point-particle
calculation.
For the second process, photon decay into an electron-positron pair
in the theory \eqref{modQED} with $\k <0$,
the point-particle calculation still holds and we have given more
details than available in the literature up till now, in particular
the momentum transfer $Q^2$ and the photon decay length.
The upshot is that the previous two-sided bound \eqref{k_bound}
from Ref.~\cite{KS2008} remains unchanged.

The multimessenger astrophysics program studying high-energy phenomena
with ultra-high-energy cosmic rays~\cite{GagnonMoore,KlinkhamerRisse1,KlinkhamerRisse2}, cosmic gamma rays~\cite{Daniel2015}, and cosmic neutrinos~\cite{DKM2014}
has developed the field of astroparticle physics over the last years and
now serves as a powerful tool to test fundamental physics symmetries.

\section*{\hspace*{-4.5mm}ACKNOWLEDGMENTS}
\vspace*{-0mm}\noindent
It is a pleasure to thank M. Risse for discussions
and the referee for useful comments.
The start-up phase of this work was supported in part
by the ``Helmholtz Alliance for Astroparticle Physics HAP,''
funded by the Initiative and Networking Fund of the Helmholtz Association,
and the main phase was supported in part
by the German Research Foundation (DFG)
under Grant No. \mbox{KL 1103/4-1.}

\begin{appendix}

\section{Photon decay into a proton-antiproton pair}
\label{app:Photon-decay-proton-antiproton-pair}

A direct parton-model calculation of the Lorentz-violating
decay process $\widetilde{\gamma} \to p+\overline{p}$
appears to be difficult if we wish to use standard fragmentation
functions~\cite{PDG2014}. Instead, we will take a
different approach employing electromagnetic form factors
(see, e.g., Ref.~\cite{Pacetti-etal2015} for a review).

The standard pointlike Dirac interaction term from \eqref{L-Dirac}
is to be generalized by the introduction of form factors
$F_{n}$. Going to momentum space, the interaction term between
protons ($p$), antiprotons ($\overline{p}$), and modified photons ($\widetilde{\gamma}$) then becomes
\bea\label{L-Dirac-formfactors}
\hspace*{-5mm}
\mathcal{L}_\text{proton}^\text{(int)} &=&
\overline\psi(p^\prime)\,
 e\,A_\mu(q) \Big[ \gamma^\mu\,F_{1}(q^2)
  + i\sigma^{\mu\nu}q_\nu/(2M) \,F_{2}(q^2)
  \nonumber\\&&
  + \ldots \big]
\psi(p)\; \delta^{4}(p^\prime-q-p)\, .
\eea
The ellipsis allows for the possibility
that this Lorentz-violating theory needs further terms.

As a first approximation, we will only use the $F_{1}$ form factor,
\bsubeqs\label{F1-tilde-Fn-tilde}
\bea\label{F1-tilde}
F_{1}&=&\widetilde{F}\,,\\[2mm]
F_{n}&=&0\,, \;\;\text{for}\;\;n \geq 2\,.
\eea
\esubeqs
In the $\k >0$ theory \eqref{modQED} with
form-factor interactions \eqref{L-Dirac-formfactors}
and \textit{Ans\"{a}tze} \eqref{F1-tilde-Fn-tilde},
the Cherenkov power becomes
\beq\label{P-tilde}
\widetilde{P}^\text{\,VCh}(E) =
 \int_{0}^{\omega_\text{max}}d\omega\;\omega \;
 \left|\widetilde{F}\left(-\frac{2\,\k\,\omega^2}{1-\k}\right)\right|^{2} \; \frac{d\widehat{\Gamma}^\text{\,VCh}}{d\omega}\, ,
\eeq
with $\omega_\text{max}$ given by \eqref{w_max} for $x=1$
and $d\widehat{\Gamma}^\text{\,VCh}/d\omega$ given by
the right-hand side of \eqref{VCh-emission-rate} for $x=1$ and $e_i^2=1$.
Here and in the following, the tilde on a quantity denotes the
use of the form factors \eqref{F1-tilde-Fn-tilde}.
Without form factor $F_2$, the expression for photon decay
in the $\k <0$ theory is relatively simple,
\bea\label{Gamma-tilde-PhD}
\widetilde{\Gamma}^\text{\,PhD}(\omega)
&=&
\left|\widetilde{F}\left(-\frac{2\,\k\,\omega^2}{1-\k}\right)\right|^{2}\;
\widehat{\Gamma}^\text{\,PhD}(\omega)\,,
\qquad
\eea
where $\widehat{\Gamma}^\text{\,PhD}$ is given by \eqref{Gamma-hat-PhD}.

\begin{figure}
\centering
\includegraphics[width=0.25\textwidth]{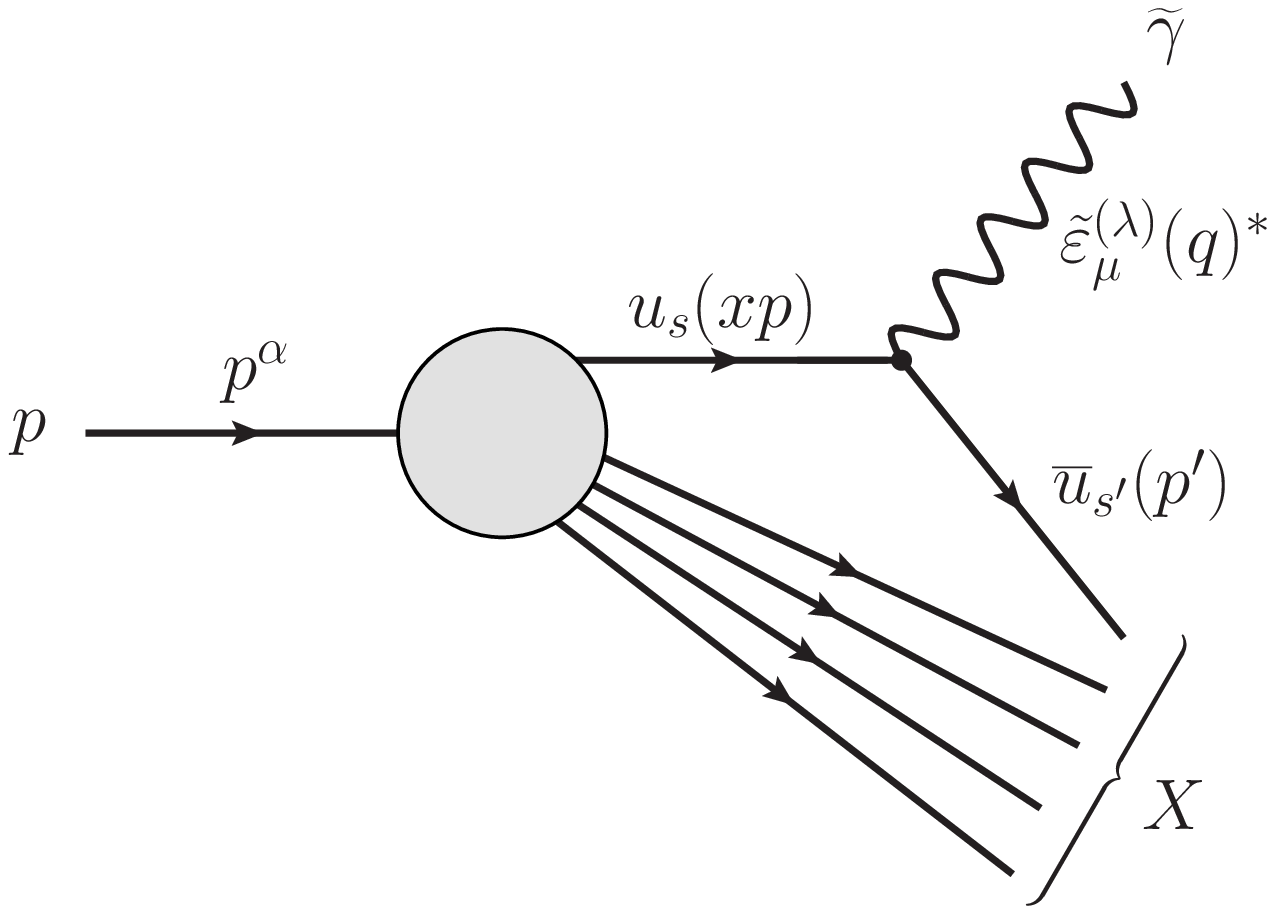}
\vspace*{-0.3cm}
\caption{Tree-level Feynman diagram for vacuum Cherenkov radiation
by a charged parton inside a high-energy proton $p$.
Lorentz-violating effects are contained in the modified
polarization vector of the outgoing photon $\widetilde{\gamma}$
and its modified dispersion relation \eqref{disp-rel}.
For energies just above threshold, there is elastic photon emission,
$p \to p+\widetilde{\gamma}$.
But, for energies significantly above threshold, also inelastic
photon emission processes occur, with additionally produced hadrons.
The present article considers the inclusive process,
$p \to X+\widetilde{\gamma}$.
}
\label{fig:FD-VCh}
\vspace*{0.05cm}
\centering           
\includegraphics[width=0.40\textwidth]{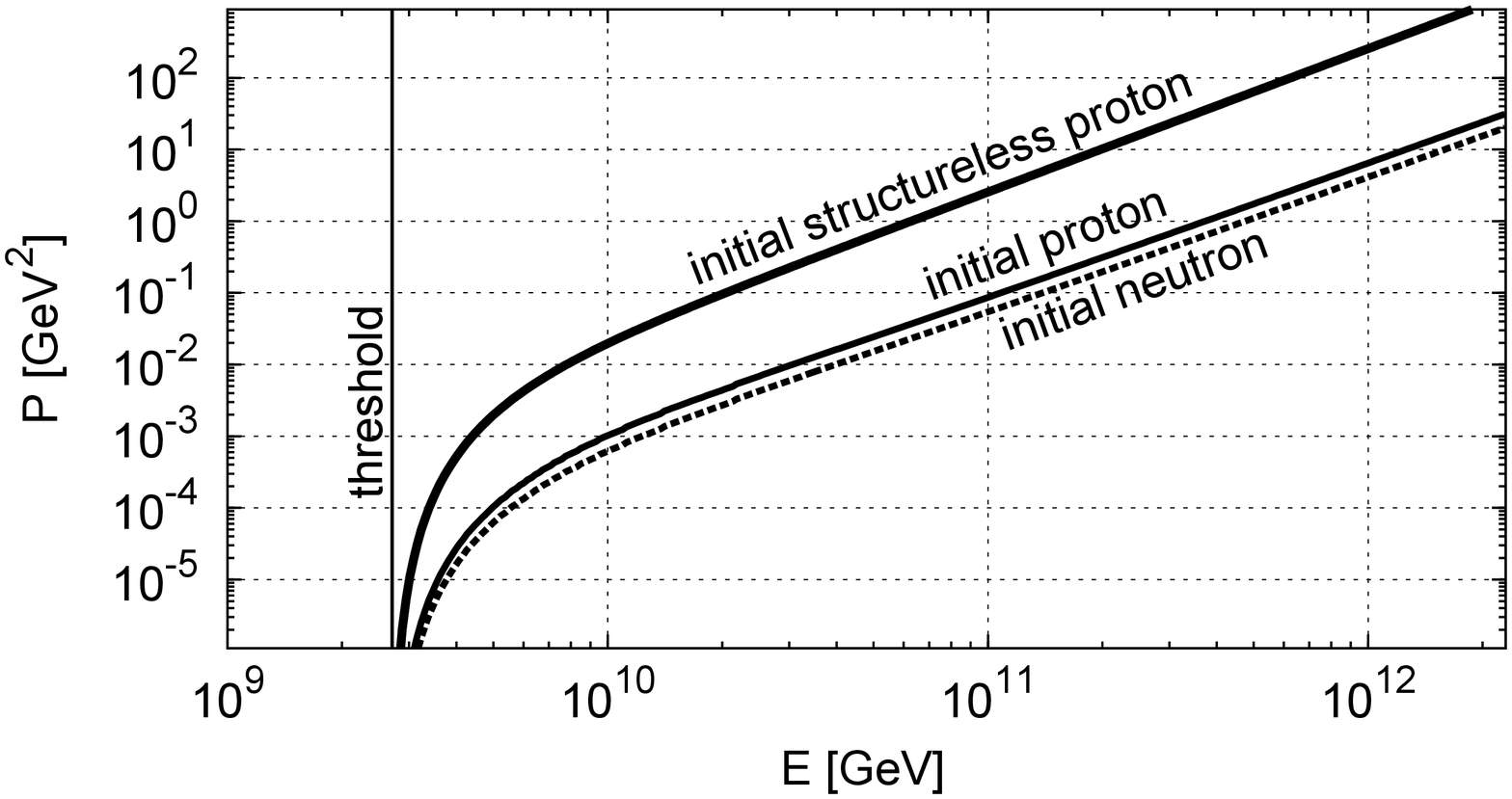}
\vspace*{-0.3cm}
\caption{Cherenkov power $P$
for the inclusive decay process $h \to X+\widetilde{\gamma}$
as a function of the energy $E$ of the initial hadron $h$, which
can be a proton, a neutron, or a structureless charged Dirac fermion
(the initial structureless fermion remains unchanged
by the photon emission).
The proton and neutron results are obtained from the parton-model
expression \eqref{P(E)} and further details are given in the main text.
The structureless-proton result follows from \eqref{P-hat(E)}.
The Lorentz-violating parameter is $\k=6\times10^{-20}$ and
the threshold energy for the proton case is given by Eq. \eqref{E_th}. }
\label{fig:VCh-Power}
\vspace*{0.05cm}
\centering          
\includegraphics[width=0.40\textwidth]{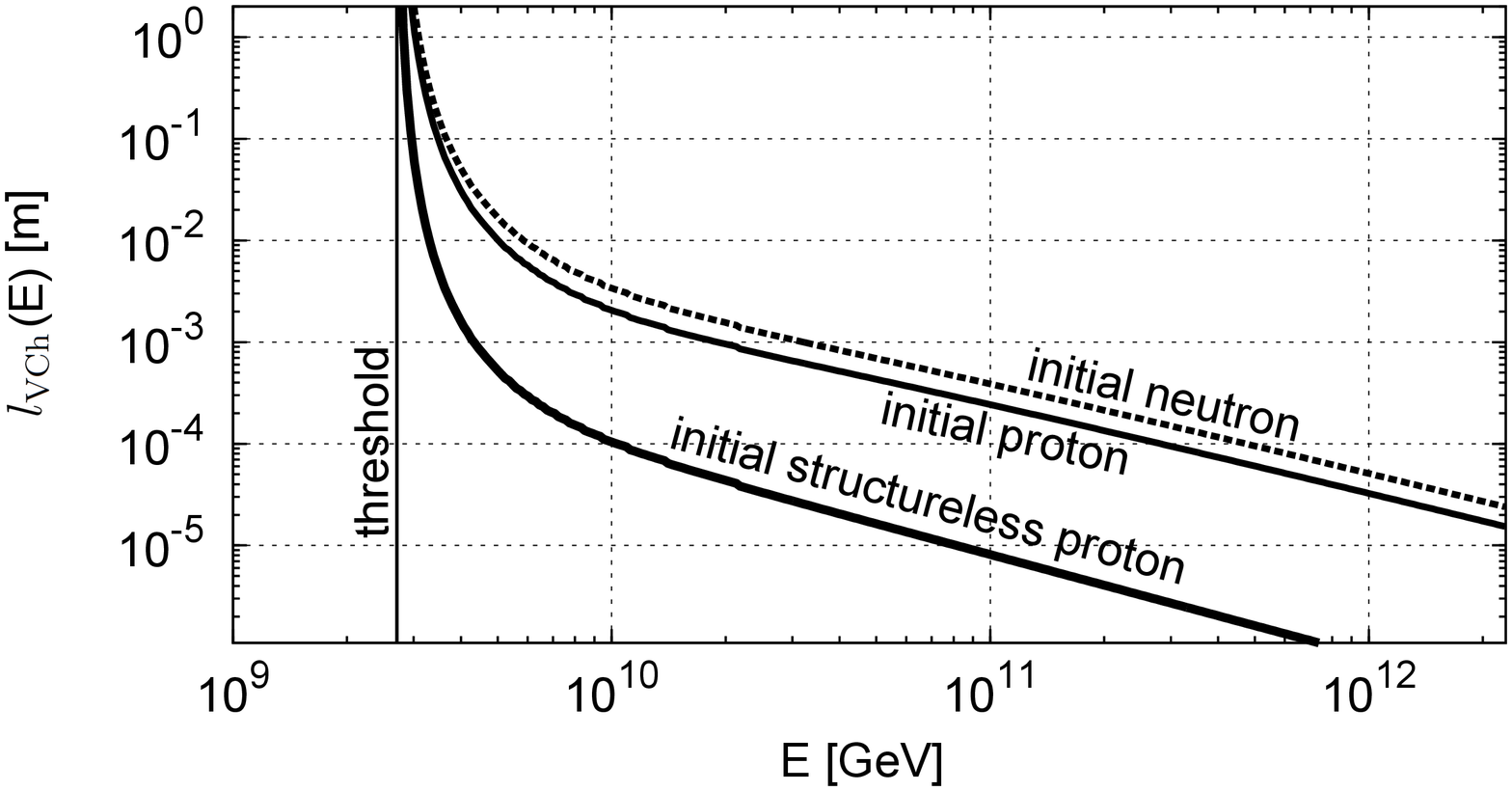}
\vspace*{-0.3cm}
\caption{Cherenkov radiation length $l_\text{\,VCh}$ in meters for a proton, a neutron, and a structureless charged Dirac fermion with $\k=6\times10^{-20}$
as in Fig.~\ref{fig:VCh-Power}. The radiation length $l_\text{\,VCh}$
is defined by \eqref{l-VCh}.}
\label{fig:VCh-radiation-length}
\vspace*{0.05cm}
\centering            
\includegraphics[width=0.40\textwidth]{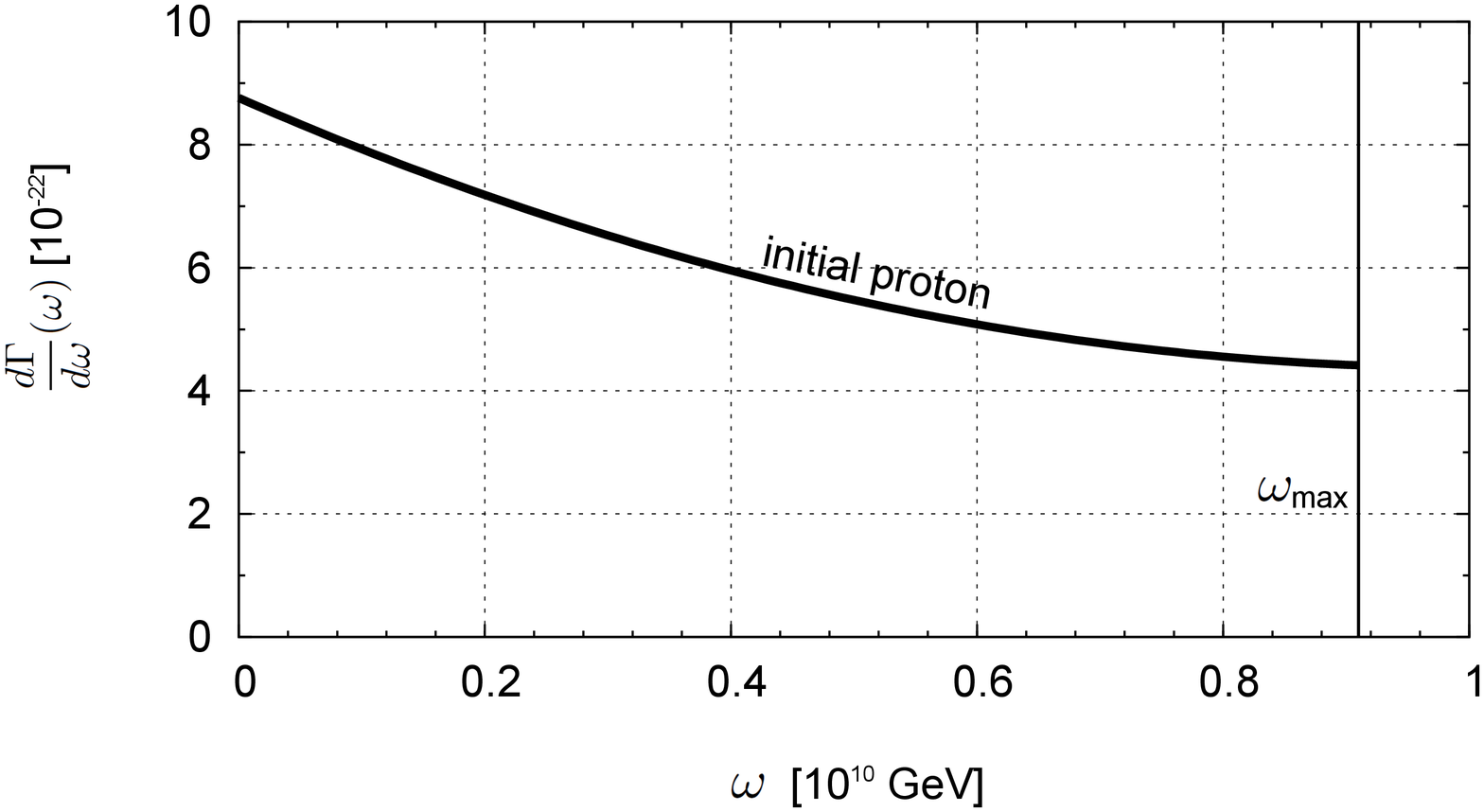}
\vspace*{-0.3cm}
\caption{Differential photon spectrum $d\Gamma/d\omega$ from vacuum
Cherenkov radiation by a proton of energy $E=10^{10}\;\text{GeV}$
in the $\k=6\times10^{-20}$ theory, as in Fig.~\ref{fig:VCh-Power}.
The differential photon spectrum is obtained from
the parton-model expression
\eqref{P(E)} by omitting the $\omega$ factor in the integrand and
not performing the integral over $\omega$.
}
\label{fig:VCh-photon-spectrum}
\end{figure}

\begin{figure}
\centering
\includegraphics[width=0.25\textwidth]{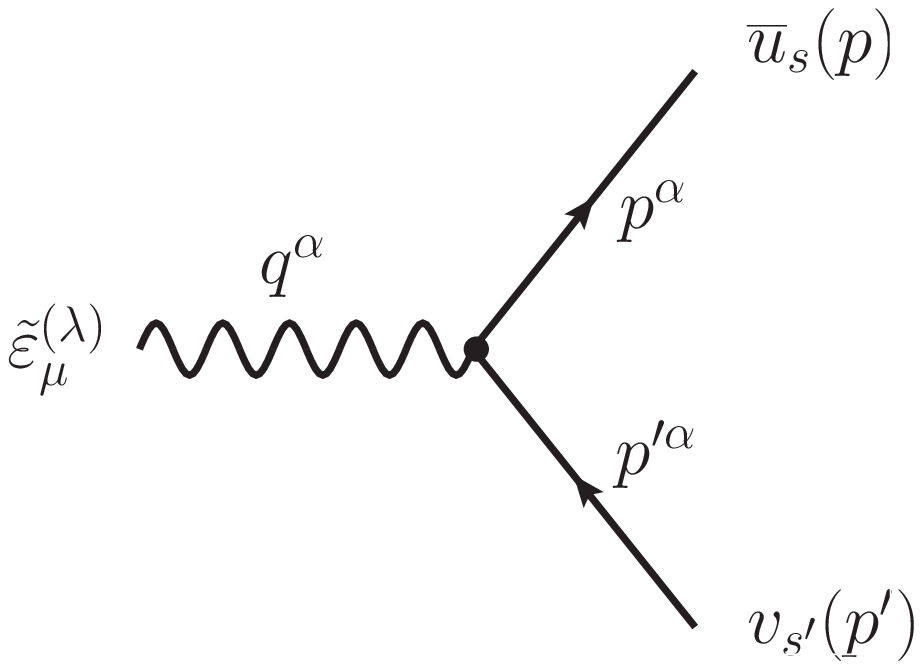}
\vspace*{-0.2cm}
\caption{Tree-level Feynman diagram for photon decay into an electron-positron pair, $\widetilde{\gamma} \to e^{-}+e^{+}$.
Lorentz-violating effects are contained in the modified polarization vector
and dispersion relation of the incoming photon.
}
\label{fig:FD-PhD}
\vspace*{.2cm}
\centering
\includegraphics[width=0.40\textwidth]{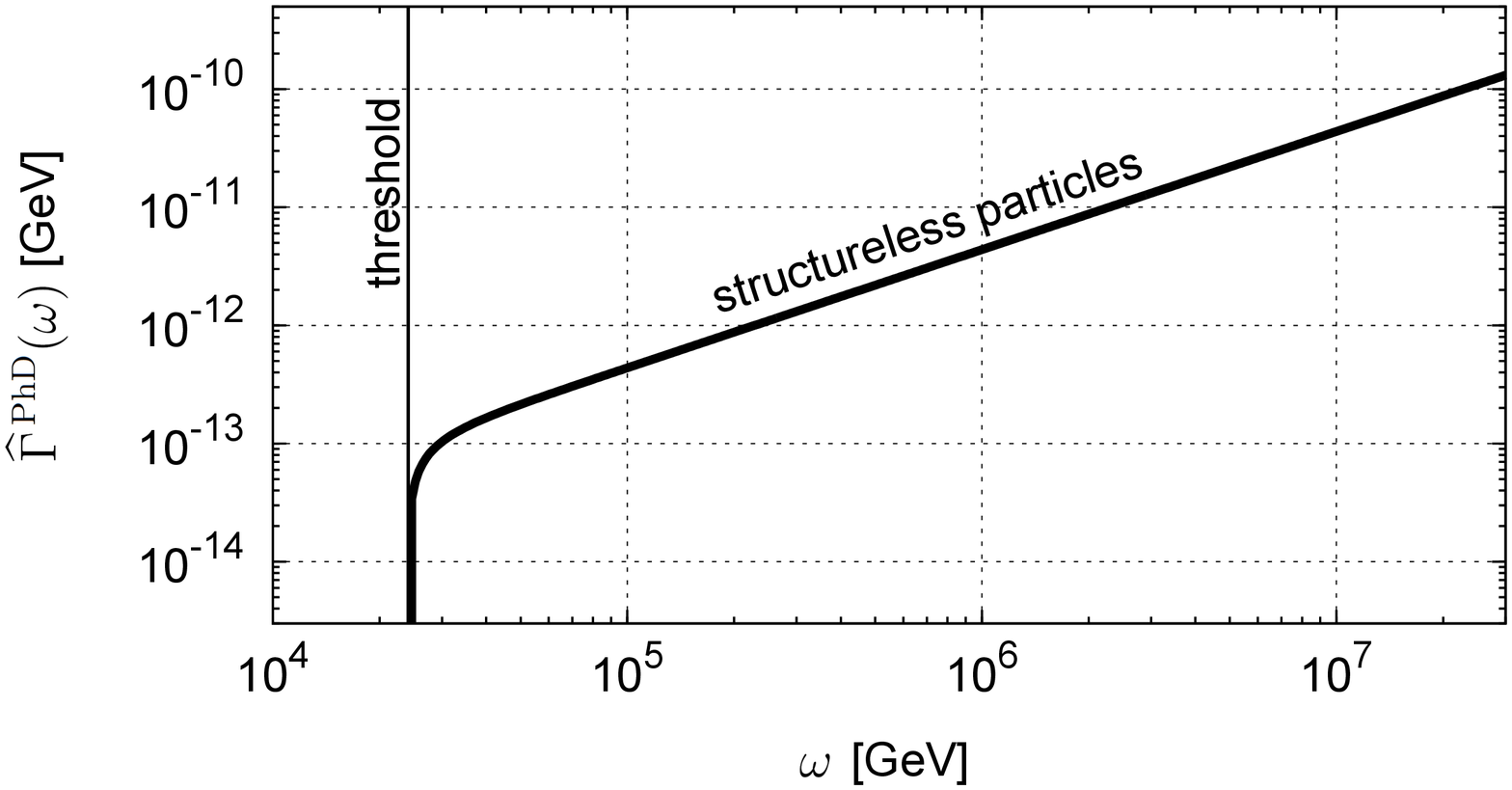}
\vspace*{-0.2cm}
\caption{Photon decay constant $\widehat{\Gamma}$ for
$\widetilde{\gamma} \to e^{-}+e^{+}$
as a function of the photon energy $\omega$.
The point-particle decay constant $\widehat{\Gamma}$ is given
by \eqref{Gamma-hat-PhD}.
The Lorentz-violating parameter is $\k=-9\times10^{-16}$ and
the threshold energy is given by Eq. \eqref{w_th}
in terms of the electron mass. }
\label{fig:PhD-Gamma}
\vspace*{.2cm}
\centering
\includegraphics[width=0.40\textwidth]{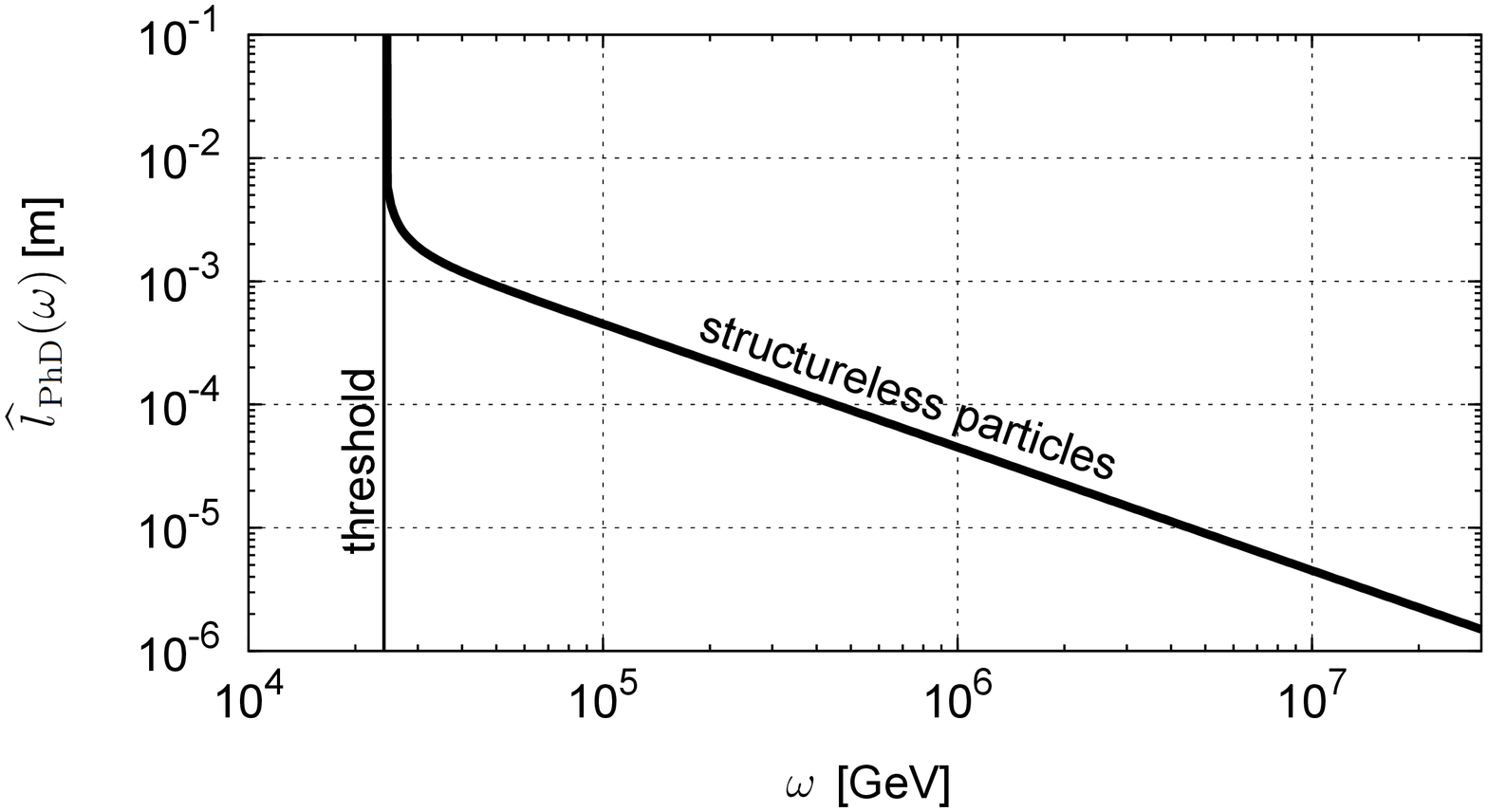}
\vspace*{-0.2cm}
\caption{Photon decay length $\widehat{l}_\text{\,PhD}$
in meters for $\widetilde{\gamma} \to e^{-}+e^{+}$ with $\k=-9\times10^{-16}$ as in Fig.~\ref{fig:PhD-Gamma}.
The decay length $\widehat{l}_\text{\,PhD}$ is defined by \eqref{l-hat-PhD}.}
\label{fig:PhD-decay-length}
\vspace*{.2cm}
\centering
\includegraphics[width=0.40\textwidth]{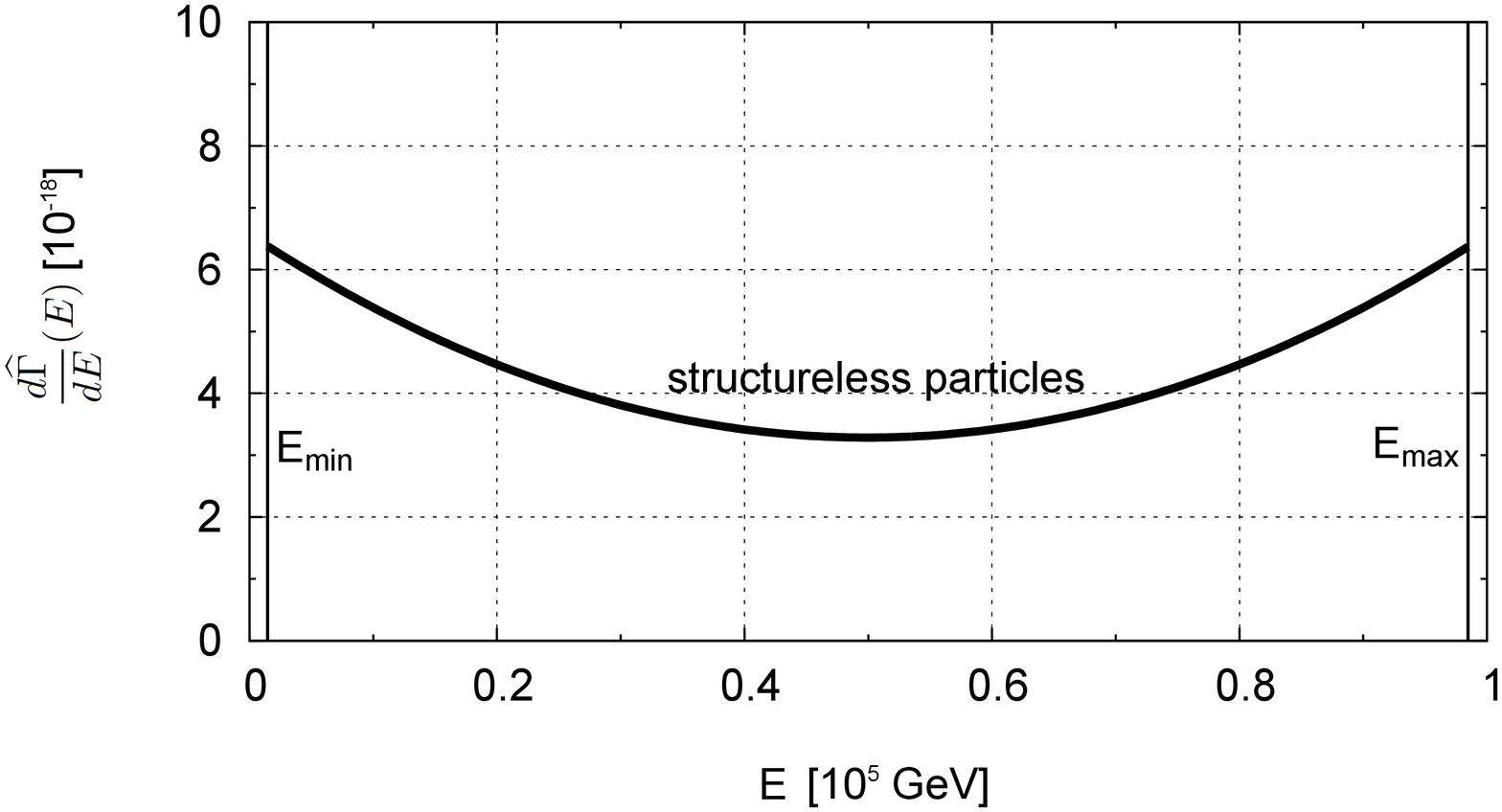}
\vspace*{-0.2cm}
\caption{
Differential electron spectrum $d\widehat{\Gamma}/dE$ from
the Lorentz-violating decay of a photon $\widetilde{\gamma}$ of energy $\omega=10^{5}\;\text{GeV}$ into an electron-positron
pair, calculated for pointlike particles in the
$\k=-9\times10^{-16}$ theory, as in Fig.~\ref{fig:PhD-Gamma}.}
\label{fig:PhD-electron-spectrum}
\end{figure}

\begin{figure}[t]
\centering
\includegraphics[width=0.40\textwidth]{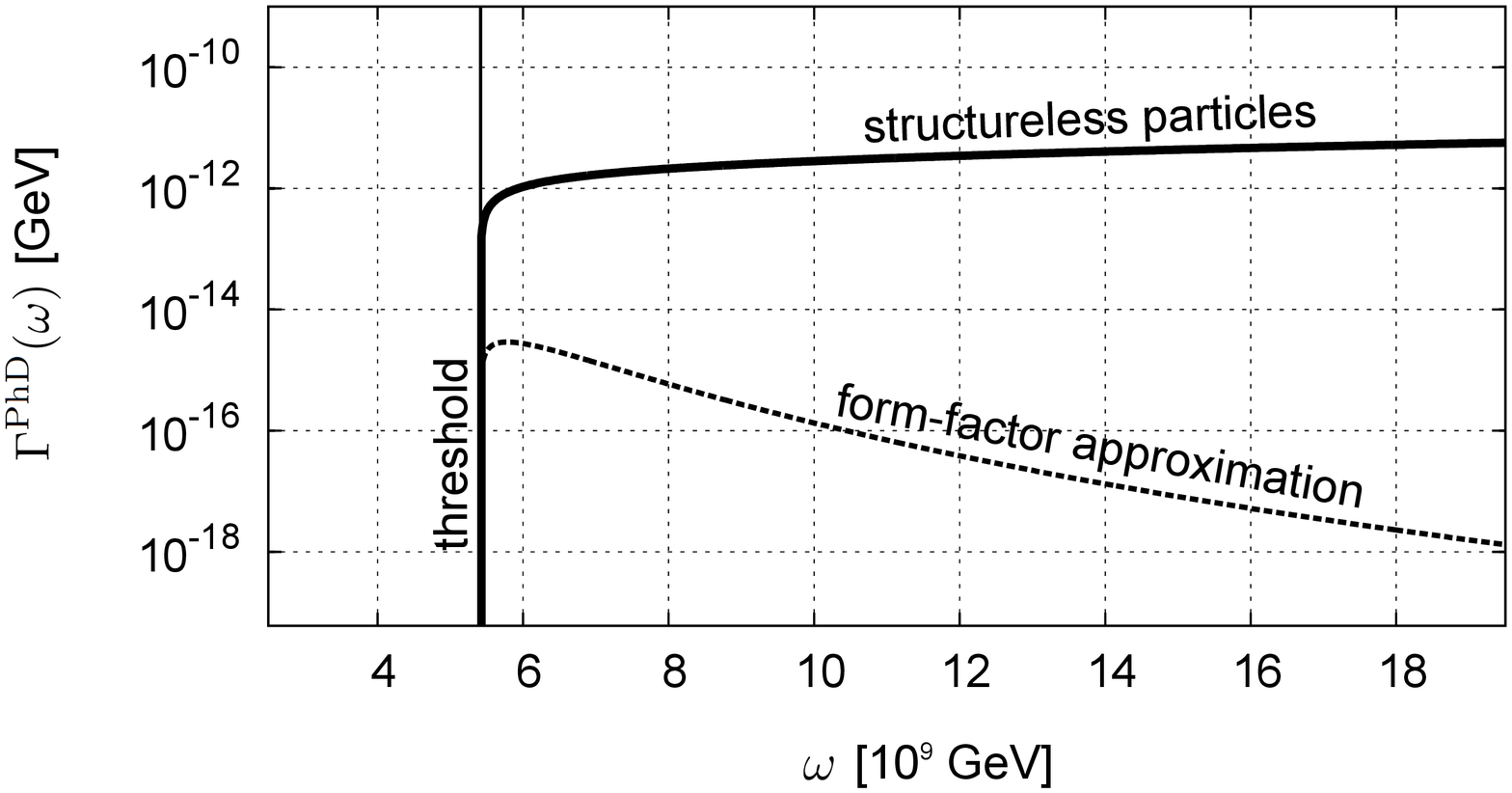}
\vspace*{-0.2cm}
\caption{Photon decay constant $\Gamma^\text{\,PhD}$ for
$\widetilde{\gamma} \to p+\overline{p}$
as a function of the photon energy $\omega$.
Shown are the results from a structureless description
and from a form-factor approximation
\eqref{L-Dirac-formfactors}--\eqref{F1-tilde-Fn-tilde} based on
the function \eqref{F1-tilde-approximation}.
The Lorentz-violating parameter is $\k=-6\times10^{-20}$ and
the threshold energy is given by Eq. \eqref{w_th} in terms of the
proton mass. }
\label{fig:PhD-Gamma-tilde}
\vspace*{15cm}
\end{figure}

For the theory with $\k=6\times10^{-20}$,
the direct parton-model results of the Cherenkov power $P$ as
shown in Fig.~\ref{fig:VCh-Power} are reproduced by inserting
$\widetilde{F}\sim \sqrt{1/20}$ into the integrand of \eqref{P-tilde}.
For timelike $q^2 >0$, relevant to photon decay in the $\k <0$ theory,
the following simple approximation can be used:
\beq\label{F1-tilde-approximation}
\widetilde{F}(q^2)
\sim \frac{\text{GeV}^4}{\big(q^2\big)^2+\text{GeV}^4}\,\;,
\eeq
whose physics motivation will be given shortly.
For the theory with $\k=-6\times10^{-20}$,
the explicit function \eqref{F1-tilde-approximation} inserted in
\eqref{Gamma-tilde-PhD} then gives the results shown in
Fig.~\ref{fig:PhD-Gamma-tilde}.
Incidentally, the same function \eqref{F1-tilde-approximation}
for spacelike $q^2<0$
inserted into the integrand of \eqref{P-tilde} in the $\k >0$ theory
gives a constant value for $\widetilde{P}^\text{\,VCh}(E)$
as $E\to\infty$.
For the parameter values of Fig.~\ref{fig:VCh-Power}, this
asymptotic $\widetilde{P}^\text{\,VCh}$
value is approximately $3\times 10^{-3}\;\text{GeV}^2$,
but note that this form-factor result
only concerns elastic photon emission, whereas the
initial-proton parton-model curve in Fig.~\ref{fig:VCh-Power}
refers to the inclusive process.

The promised background on the choice \eqref{F1-tilde-approximation}
is as follows. In terms of Sachs form factors
$G_{E}(q^2) \equiv  F_1(q^2) + [q^2/(4M^2)] F_2(q^2)$
and $G_{M}(q^2) \equiv F_1(q^2) +  F_2(q^2)$,
a common working hypothesis is $|G_{E}|=|G_{M}|=G_\text{eff}$.
Considering real form factors (appropriate for $|q^2| \to\infty$),
this hypothesis
corresponds precisely to our assumption \eqref{F1-tilde-Fn-tilde},
with $F_1=G_\text{eff}$ and $F_2=0$, so that \eqref{Gamma-tilde-PhD}
can be read as $|G_\text{eff}|^2\times \widehat{\Gamma}^\text{\,PhD}$.
The experimental data for $G_\text{eff}(q^2)$ with
timelike $q^2$ between approximately $4\;\text{GeV}^2$ and
$30\;\text{GeV}^2$ (see Fig.~10 of Ref.~\cite{Pacetti-etal2015})
is very roughly and qualitatively fitted by the simple
function \eqref{F1-tilde-approximation} with the expected
asymptotic behavior for $|q^2| \to\infty$,
leaving aside the precise behavior at the physical
threshold $(q^2)_\text{phys}=4M^2 \approx 3.53\;\text{GeV}^2$.

Returning to the photon-decay process
of Fig.~\ref{fig:PhD-Gamma-tilde},
the question is where the definitive curve for
a realistic proton lies. We expect the definitive curve
for the inclusive process $\widetilde{\gamma} \to p+\overline{p}+X$
to lie closer to the structureless curve than
to the form-factor curve.
Heuristically we can give the following argument.
In the modified QED model \eqref{modQED}
with form factors \eqref{L-Dirac-formfactors},
on the one hand, the photon $\widetilde{\gamma}$
directly creates a complete extended proton particle and
a complete extended antiproton particle.
In the parton-model version, on the other hand, the
photon $\widetilde{\gamma}$ first creates a pointlike
antiquark and a pointlike quark, and then dresses them
by pulling (anti)quarks and gluons out of the vacuum,
possibly creating more hadrons than a single
proton-antiproton pair. For relatively large momentum transfer
($q^2 \gtrsim \text{GeV}^2$),
it seems that the parton-model description is more appropriate,
but the definitive calculation remains to be performed.
It may also be instructive to compare to the
case of $p\overline{p}$ production in electron--positron annihilation,
for which many experimental results exist
(cf. Sec.~20.6 in Ref.~\cite{PDG2014}).

\end{appendix}

\newpage

\newpage

\newpage

\end{document}